\documentclass{aa}  

\usepackage{multirow}
\usepackage{graphicx}
\usepackage[font=small,skip=0pt]{caption}
\usepackage{txfonts}

\begin{document} 
	\titlerunning{NuSTAR observations of heavily obscured  BAT  AGN}
\authorrunning{Georgantopoulos et al.}

   \title{NuSTAR observations of heavily obscured Swift/BAT  AGN: constraints on the Compton-thick AGN fraction}

   \author{I. Georgantopoulos\inst{1}, A. Akylas\inst{1}}


   \institute{IAASARS, National Observatory of Athens, I.  Metaxa \& V.  Pavlou, 15236,  Greece\\
              \email{ig@noa.gr}
             }


 
  \abstract
 { The evolution of the accretion history of the Universe has been studied in unprecedented detail owing to recent X-ray surveys performed 
 by {\it Chandra} and {\it XMM-Newton}.
  A missing piece of information is the most heavily obscured or Compton-thick  AGN which evade 
 detection even in  X-ray surveys owing to their extreme hydrogen column densities which exceed  $\rm 10^{24} cm^{-2}$. 
 Recently, the all-sky hard X-ray survey performed by {\it Swift}/BAT brought a breakthrough  allowing the detection of many of these 
 AGN. This is because of its very high energy bandpass  ($\rm 14-195$ keV) which helps to minimise 
 attenuation effects. In our previous work, we have identified more than  50 candidate Compton-thick AGN in the local Universe, 
 corresponding to an {\it observed} fraction of about 7\% of the total AGN population.  This number can be converted to the {\it intrinsic}
 Compton-thick AGN number density,  only if we know  their  exact selection function. This  function sensitively depends 
 on the form of the Compton-thick AGN  spectrum, that is the energy of their absorption turnover,
 photon-index and its cut-off energy at high energies, as well as the strength of the reflection component on the matter surrounding the nucleus. 
 For example, the reflection component at hard energies 20-40 keV antagonises the number density of missing Compton-thick AGN 
 in the sense that the stronger the reflection the easier these sources are detected in the BAT band. 
 In order to constrain  their number density, 
 we analyse the spectra of 19 Compton-thick AGN which have been detected with {\it Swift}/BAT and have been subsequently 
 observed with {\it NuSTAR} in the 3-80 keV band. We analyse their X-ray spectra using the {\sc MYTORUS} models of Murphy and Yaqoob  which properly take into
  account the Compton scattering effects. These are combined with physically motivated Comptonisation models which accurately describe the 
  primary coronal X-ray emission. 
 We derive absorbing column densities which are consistent with those derived by the previous {\it Swift}/BAT analyses. 
 We estimate the coronal temperatures to be roughly between 25 and 80 keV corresponding to high energy cut-offs
 roughly between 75 and 250 keV. Furthermore, we find that the majority of our AGN lacks a strong reflection component 
 in the 20-40 keV band placing tighter constraints on the intrinsic Compton-thick 
 AGN fraction.  Combining these results with our  X-ray background synthesis models, we estimate a  Compton-thick AGN fraction
  in the local Universe of $\approx20\pm3$ \% relative to the type-II AGN population .
 }


   \maketitle

%

\section{Introduction}

   The vast majority of galaxies contain a supermassive black hole in their centres \citep
   [e.g.][]{ferrarese2000}. A fraction of these emit copious amounts of radiation as  in-falling  
   matter forms an accretion disk around the black hole. These Active Galactic Nuclei (AGN) constitute at least half of the galaxy population \citep[]{ho1997a, ho1997b}. 
   The majority of AGN  are covered by veils of dust and gas (e.g. Ricci et al. 2017) which probably have a toroidal form \citep{wada2009}.  
   In most cases this obscuring screen can obliterate the optical emission and reradiate in the IR. X-rays can more easily penetrate these obscuring screens. 
   However, in the most extreme obscuration even the X-ray radiation can be blocked. 
   These are the Compton-thick AGN 
   having column densities exceeding $\rm 10^{24}$ $\rm cm^{-2}$. In this case,  the X-rays are Compton scattered on the free electrons of the obscuring screen. 
   As these heavily obscured AGN represent the evasive side of the accretion history of the Universe, it is of primary importance to detect them. 
   However, the methods employed to constrain their number appear to give diverging results. 

The most commonly used method is that of X-ray background synthesis models. The spectrum of the X-ray background,  that is the integrated X-ray light 
produced by all X-ray sources in the history of the Universe, presents a peak at 20-30 keV as measured by the {\it HEAO-1, XMM-Newton, BeppoSAX} and the {\it INTEGRAL} 
missions \citep[e.g.][]{gruber1999, frontera2007, churazov2007, moretti2009}. X-ray background 
synthesis models attempt to reconstruct the spectrum of the X-ray background using the AGN luminosity function, together with a model for  the AGN X-ray  spectrum. 
These models have been pioneered by \cite{comastri1995} and were later refined by \cite{gilli2007}, \cite{treister2009}, \cite{draper2009},  \cite{akylas2012}, \cite{vasudevan2013}, \cite{ueda2014}, \cite{esposito2016}. 
In order to reproduce the hump of the X-ray background, a significant number of Compton-thick AGN is necessary. However, their fraction remains uncertain 
with figures ranging roughly  between 10 and 50 \%. This discrepancy is attributed mainly to the fact that there is a degeneracy between the required fraction of 
Compton-thick AGN and the strength of the total AGN population reflection component. This reflection component originates from scattering and 
subsequent absorption of the X-ray radiation on surrounding cold material \citep[e.g.][]{george1991} and results to the  flattening of the observed AGN spectrum. 
The higher the AGN population reflection component the lower the number of Compton-thick AGN required. 
Interestingly enough, the reflection component strength does not refer to  the Compton-thick AGN but to the average reflection component of the total AGN population.

Recently, attention has shifted to the direct detection of Compton-thick  AGN using X-ray spectroscopy. The sensitive observations performed with {\it Chandra}
and {\it XMM-Newton} allowed the detection of many faint candidate Compton-thick AGN in the 2-10 keV band mainly in the Chandra
 Deep Field South area \citep[e.g.][]{georgantopoulos2013, brightman2014, buchner2015} 
 resulting in the derivation of  the Compton-thick AGN luminosity function and its evolution. 
 Additional studies have focused on the ultra-hard band above 10 keV 
where the X-ray attenuation is less severe. The {\it Niel Gehrels Swift}, {\it INTEGRAL} and {\it NuSTAR} missions which operate in this band, readily provide this possibility. 
 The BAT hard X-ray instrument onboard {\it Niel Gehrels Swift} is continuously  performing an all sky survey in the $14-195$ keV band. 
 As it does not carry imaging X-ray optics it can  probe flux limits of 
only a few times $10^{-12}$ $\rm erg~cm^{-2}~s^{-1}$. \cite{ricci2015} and \cite{akylas2016} analysed the X-ray spectra of 688 candidate AGN in the 
70-month BAT all-sky survey \citep{baumgartner2013}. Fitting the X-ray spectra using Bayesian statistics, \cite{akylas2016} find about 
50 Compton-thick  AGN candidates in overall agreement with the results of \cite{ricci2015}. The observed fraction of Compton-thick AGN amounts 
to about 7\% of the total AGN population. This number can be extrapolated to the intrinsic fraction of Compton-thick AGN, only by knowing the exact X-ray 
spectrum of Compton-thick AGN  i.e. their selection function. \cite{akylas2016} find that the intrinsic fraction of Compton-thick AGN, in the local Universe, is between 10 and 30\%
 assuming a reflection component which contributes between 0 and 5\% to the total 2-10 keV flux. These estimates are in good agreement with the estimates 
 of \cite{ricci2015}, \cite{burlon2011} and \cite{malizia2009}. The advantage of the above  direct methods is that only knowledge of the X-ray spectrum of Compton-thick AGN is required instead of the 
 average AGN population.

In this paper, we present an X-ray spectral analysis of the Compton-thick AGN in \cite{akylas2016} which have publicly available {\it NuSTAR} spectra. 
The imaging X-ray optics of NuSTAR  \citep[]{harrison2013} allow the derivation of excellent quality spectra. 
These are used to constrain with the highest accuracy yet, the fraction of Compton-thick AGN in the local Universe. 
We adopt
$\rm H_o=75\,km\,s^{-1}\,Mpc^{-1}$, $\rm\Omega_{M}=0.3$, and $\Omega_\Lambda=0.7$
throughout the paper. 

\section{The X-ray Sample \& data}
\subsection{BAT}
The {\it Niel Gehrels Swift} gamma-ray burst (GRB) observatory \citep{gehrels2004} was launched in November 2004 and has been
continually observing the hard X-ray ($\rm 14-195$ keV) sky with the Burst Alert Telescope (BAT). 
 BAT  is a large, coded-mask
telescope optimised to detect transient GRBs and is
designed with a very wide field-of-view of $\sim$ 60$\times$100 degrees. {\it Niel Gehrel Swift}'s 
observation strategy is to observe targets with the
narrow field-of-view X-ray telescope (XRT), in the $\rm 0.3-10$ keV band until a new GRB 
is discovered by BAT, at which time
{\it Niel Gehrel Swift} automatically slews to the new GRB to follow up with the
narrow field instruments until the X-ray afterglow is below the XRT detection limit.
 \citet{baumgartner2013}  presented the catalogue of sources detected in 70 months of observations with BAT. 
 The BAT 70 month survey has detected 1171 hard X-ray sources down to a flux level of $\rm 10^{-11} erg ~cm^{-2} ~ s^{-1}$. 
 The majority of these sources are AGN.  
 
\citet{akylas2016} analysed the X-ray spectra of these sources employing Bayesian statistics. They  considerd 688 sources classified
according to the NASA/IPAC Extragalactic Database in the following types: (i) 111 galaxies, (ii) 292 Seyfert I (Sy 1.0-
1.5), (iii) 262 Seyfert II (Sy 1.7-2.0), and (iv) 23 sources of type
other AGN. Radio-loud AGN have been excluded since their
X-ray emission might be dominated by the jet component. QSOs
 have been also excluded from the analysis since the fraction of highly
absorbed sources within this population is negligible. 
 They have combined the BAT with the XRT data (0.3-10 keV)
at softer energies adopting a Bayesian approach to fit the data using Markov chains. This allows us to consider all sources as potential
Compton-thick candidates at a certain level of probability. 
 53 sources present a non-zero probability of being Compton-thick corresponding to 40 effective Compton-thick sources.
These represent 7\% of the sample in excellent agreement with the figure reported in \cite{ricci2015} and \cite{burlon2011}. 
 Out of these, 38 sources have a probability of more than 70\% being Compton-thick. 
   
\subsection{NuSTAR}
The Nuclear Spectroscopic Telescope Array, NuSTAR,  \citep{harrison2013} launched in June 2012, is the first
orbiting X-ray observatory which focuses light at high energies (E $>$ 10 keV). It consists of two co-aligned focal
plane modules (FPMs), which are identical in design. Each FPM covers
the same 12 x 12 arcmin portion of the sky, and comprises of four Cadmium-Zinc-Telluride detectors. 
NuSTAR operates between 3 and 79 keV, and
provides an improvement of at least two orders of magnitude  in sensitivity compared to
previous hard X-ray orbiting observatories, E$>$10 keV.
This is because of its excellent spatial resolution which is 58 arcsec half-power-diameter. 
The energy resolution is  0.4 and 0.9 keV at 6 and 60 keV respectively.

 All  the 38 candidate Compton-thick objects in \cite{akylas2016},  with a probability $>$70 \% have either been observed 
 or are scheduled for observation with {\it NuSTAR}. We analyse here the 19 sources which are currently publicly available. 
The details of the {\it NuSTAR} observations for each object are given in table \ref{data}. 
We extracted the spectra from each FPM using a circular aperture region of 30 arcsec -radius (corresponding to
$\sim$50\% {\it NuSTAR} encircled energy fraction, ECF. 
We processed the data  with the {\it NuSTAR} Data Analysis Software ({\sc NUSTARDAS}) v1.4.1within {\sc HEASOFT} v6.15. The
{\sc NUPIPELINE}  script was used to produce the calibrated and cleaned event files using standard filter flags.
We extracted the spectra and response files using the
{\sc NUPRODUCTS}  task. 

\begin{table*}
\tiny
\caption{The NuSTAR observations}
\centering
\setlength{\tabcolsep}{0.7mm}
\begin{tabular}{lccccc}
       \hline
Name & redshift & Gal. $\rm N_H$ & obsID& Exposure & Net counts \\
           &              &  $\rm \times 10^{20} cm^{-2}$ &  &      ksec  &   3-79keV\\
\hline
NGC1068 & 0.0038         &   2.9        &   60002030002   & 57.8 & 18900\\
Circinus    &    0.0014&   5.6 & 60002039002   & 53.9 & 87962  \\
NGC6240 &   0.0244 &   4.9  & 60002040002&  30.9 & 8716 \\
NGC4945 &   0.0019 &    13.9  &  60002051004 & 54.6 & 45725  \\
NGC424 &    0.0118 &   1.6  & 60061007002  & 15.5 & 1592   \\
2MFGC2280 & 0.0151  &  40.6    &  60061030002& 15.9 & 640  \\
NGC1194 &   0.0136 &   7.1  &  60061035002 & 31.5 & 4060  \\
MGC06-16-028& 0.0157  &   6.2 & 60061072002 & 23.6 & 1649   \\
NGC3079 &   0.0037&  0.8    & 60061097002 & 21.5 &   1470 \\
NGC3393 &  0.0125&     6.0 & 60061205002&  15.7 & 1609 \\
NGC4941 &   0.0037&    2.4  &  60061236002 & 20.7  & 1491  \\
NGC5728 &  0.0037 &    7.8  &  60061256002& 24.4 & 7344 \\
ESO137-34 &  0.0091&     24.7  &  60061272002& 18.5 & 2043   \\
NGC7212&   0.0266& 5.5    & 60061310002 &  24.6 & 1464  \\
NGC1229 & 0.03623  &    1.7  &  60061325002 & 24.9 & 1700  \\
NGC6232 &  0.0148 &    4.5  &  60061328002 &18.1 &  284    \\
2MASSXJ09235371-3141305 & 0.0424  &    12.8  & 60061339002 & 21.3 &  2940\\
NGC5643 &  0.0040 &    8.3  & 60061362002 & 22.5 &  1664   \\
NGC7130 &    0.0161 &  1.9 & 60261006002 & 42.1 &  1302  \\
\hline
\label{data}
\end{tabular}
\end{table*}

\section{X-ray spectral modelling}

Recently, sophisticated spectral models have been developed that estimate the detailed spectra of Compton-thick AGN that is they take into account 
 Compton scattering and reprocessing by Compton-thick material that surrounds the nucleus. 
 These include the models of \citet{brightman2011} and the models of \citet{yaqoob2011}. 
 Both models employ Monte Carlo simulations to account for Compton scattering and the geometry of the obscuring screen. 
 The models of \citet{brightman2011} refer to both spherical coverage and a torus geometry. These estimate self-consistently the 
 Compton scattering along the line of sight as well as the reflection from surrounding matter. 
 Instead, the toroidal Compton-thick X-ray re-processor model, {\sc MYTORUS} of \citet{murphy2009} and \citet{yaqoob2011} uses a torus 
geometry. The torus has a diameter which is characterised by the equatorial column density, $\rm N_{H},Z$. The model
assumes a configuration in which the global covering factor of the re-processor is 0.5, corresponding to a solid angle subtended
by the torus of $\rm 2\pi$. The \citet{yaqoob2011} model employs three separate components: the line of sight component, Z, 
an absorbed reflected component along the line of sight, S90, and an unabsorbed reflected component S0. 
In our spectral fits, we choose to employ the spectral models of \citet{yaqoob2011}. 
This is because this model allows for free relative normalisations between the different components in contrast to the self-consistent models of \citet{brightman2011}.
 Then the \citet{yaqoob2011} model can accommodate differences in the actual geometry  and time delays between direct, scattered and fluorescent line photons.
 We note that the reflected emission is modelled using the appropriate column density instead of  
 being modelled using reflection on a slab of infinite column density, cf. the PEXRAV model \citet{Zdziarski1995}.
   The full general geometrical set-up of this model can be visualised in Fig. 2 of \citet{yaqoob2012}.
   
 In detail,  our modelling consists of the following components:
 a) an unabsorbed  power-law originating from scattered emission on Compton-thin material, usually referred  as the 
 "soft excess" in type-2 AGN. 
 b) the direct zero-th order primary emission   absorbed by the line 
  of sight column $\rm N_{H},Z$ denoted by the multiplicative model mytorus\_Ezero .
  Following \citet{yaqoob2012}, we use  the physically motivated model {\sc COMPTT} of \citet{titarchuk1994}. This model estimates the scattering 
  of soft accretion disc photons on a hot thermal corona with temperature $\rm kT$ and optical depth $\tau$. According to this thermal comptonisation model the 
  resulting cut-off energy is $\sim$3kT.  There is  dependence between the 
  optical depth and the temperature  of the corona, in the sense that the higher the temperature of the corona the lower the value of $\tau$. In the case 
  where the optical depth cannot be constrained from the spectral fits, 
  we chose to fix the value of the optical depth to $\tau=2$. This value corresponds roughly to a photon index of $\Gamma\approx 1.9$ 
  for a coronal temperature of $\rm kT\sim 50$ keV (the average value estimated in our spectral fits below) in the case 
   of spherical geometry \citep{titarchuk1994, longair2011}. 
  $\Gamma=1.9$ is the mean photon-index value observed in 
  AGN \citep{nandra1994} and  is also the best fit photon-index found in our X-ray background synthesis models \citet{akylas2012}. 
  c) a reflected {\it absorbed} component  S90 with a column density $\rm N_{H},S90$ 
   denoted by the model mytorus\_scattered.
  d) a reflected {\it unabsorbed} component,  S0, from a column density$\rm N_H,S0$ that could possibly originate from the back-side of the torus. 
  e) An FeK$\alpha$, FeK$\beta$ and ionised Fe lines at 6.4, 7.02 and 6.7 keV respectively denoted by the model {\sc mytl} arising from 
     the line of sight absorbed scattered radiation
     f) The same lines as above denoted again by the model {\sc mytl} arising from 
     un-absorbed scattered radiation. 
 Here, we assume that all the above column densities have the same value. 
  The normalisations of the two scattered components and of the zero-th order component are allowed to vary independently.
        In {\sc XSPEC} notation, the model can be written as:
  
  \begin{eqnarray}
  \tiny
   \rm model & = & wa[1]*(po[2]+ \nonumber \\
   && comptt[3]*etable{mytorus\_Ezero}[4] + \nonumber \\
 && atable {mytorus\_scatteredkT}[5]+ \nonumber \\
  && atable {mytorus\_scatteredkT}[6] + \nonumber \\
  && atable{mytl}[7]) \nonumber \\
  && atable{mytl}[8])) \nonumber \\
  \nonumber
  \end{eqnarray}

The scattering models  mytorus\_scatteredkT and atable{mytl} correspond to a temperature kT of the corona obtaining discrete values 
between 16 and 100 keV. 
The use of these models requires  an iterative process. More specifically, in order to estimate the output reflection spectrum, it is necessary to 
 have a priori knowledge of the primary radiation spectrum and  in particular of the cut-off. In the {\sc MYTORUS} models, this is a free parameter in 
 the zero-th component spectrum but not in the reflection spectra.  In the first run we give as an input an initial cut-off energy of
 50 keV in order to estimate the reflection components S0 and S90. Then the exact cut-off energy is estimated from the fit to the zero-th order
 component. This value is fed back to the reflection components and the process is repeated. 
 In the above process we leave the {\it NuSTAR} and the BAT normalisations untied. 
    We use the {\sc XSPEC} v12.8 spectral fitting package \citep{arnaud1996}. The appropriate Galactic absorption column density has been included in all fits. 
Owing to the high number of counts we employ the $\chi^2$ statistic. We bin the spectra, using the task {\sc GRPPHA} so that they contain at least 20 counts per bin. 
All errors quoted refer to the 90\% confidence level. 

\begin{figure}
\begin{center}
\includegraphics[width=80mm]{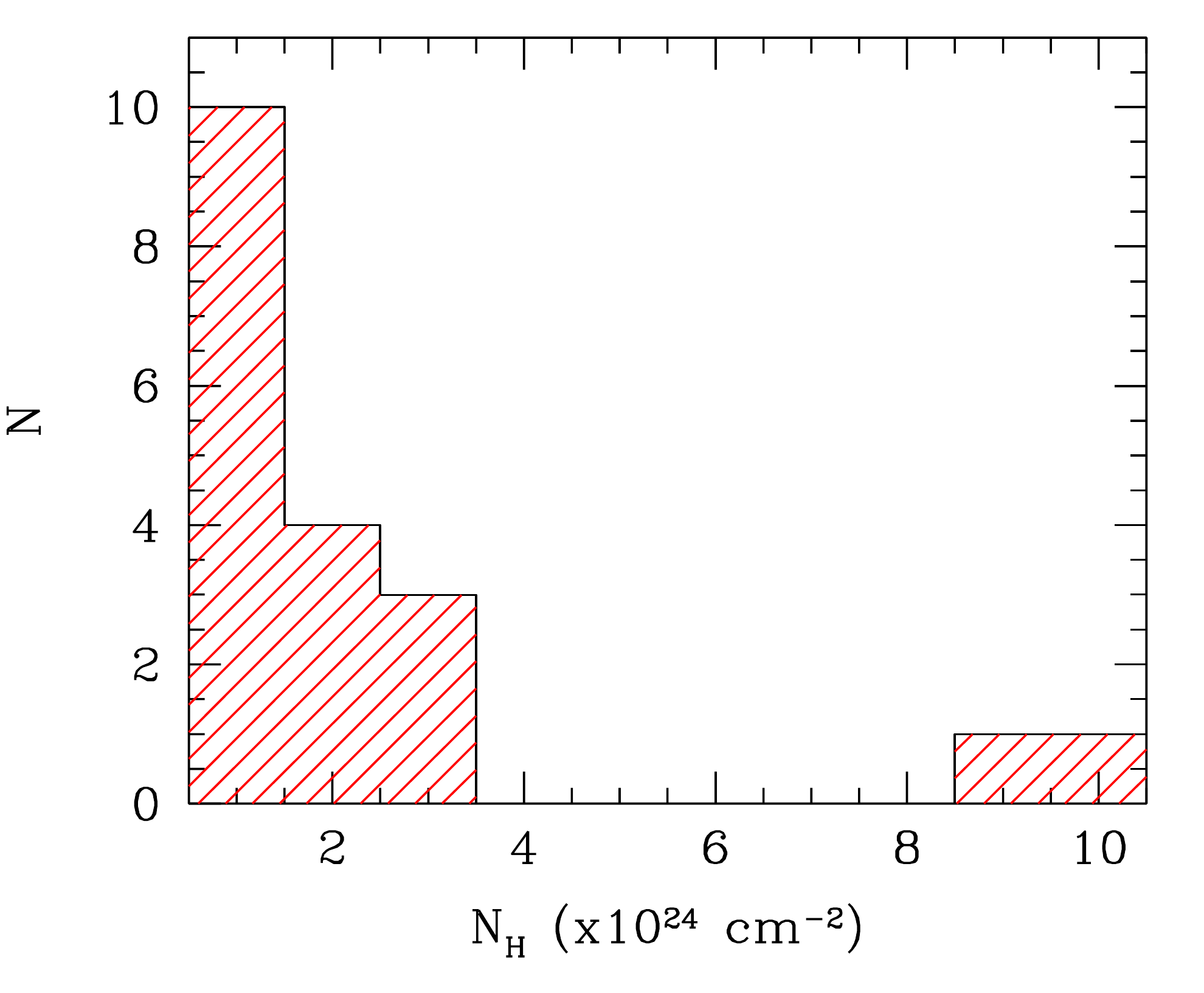}
\end{center}
\caption{The column density distribution for all 19 sources in our sample.}
\label{nh}
\end{figure}


\begin{figure}
\begin{center}
\includegraphics[width=80mm]{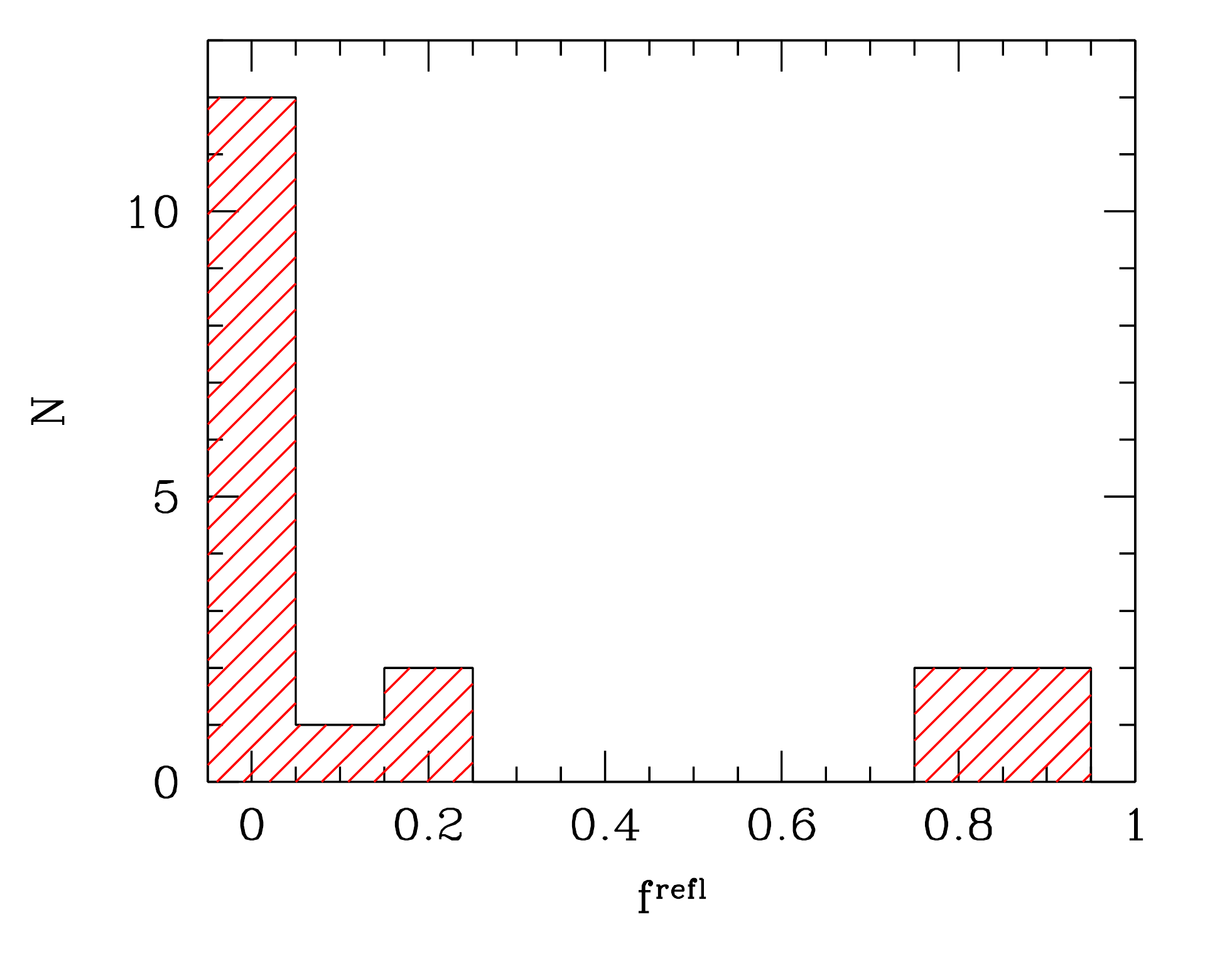}
\end{center}
\caption{The reflection fraction distribution ($f^{refl}=Flux[S0]/(Flux[Z]+Flux[S90]+Flux[S0]$ ) in the 20-40 keV band for all 19 sources  in our sample}.
\label{refl}
\end{figure}

\begin{figure}
\begin{center}
\includegraphics[width=80mm]{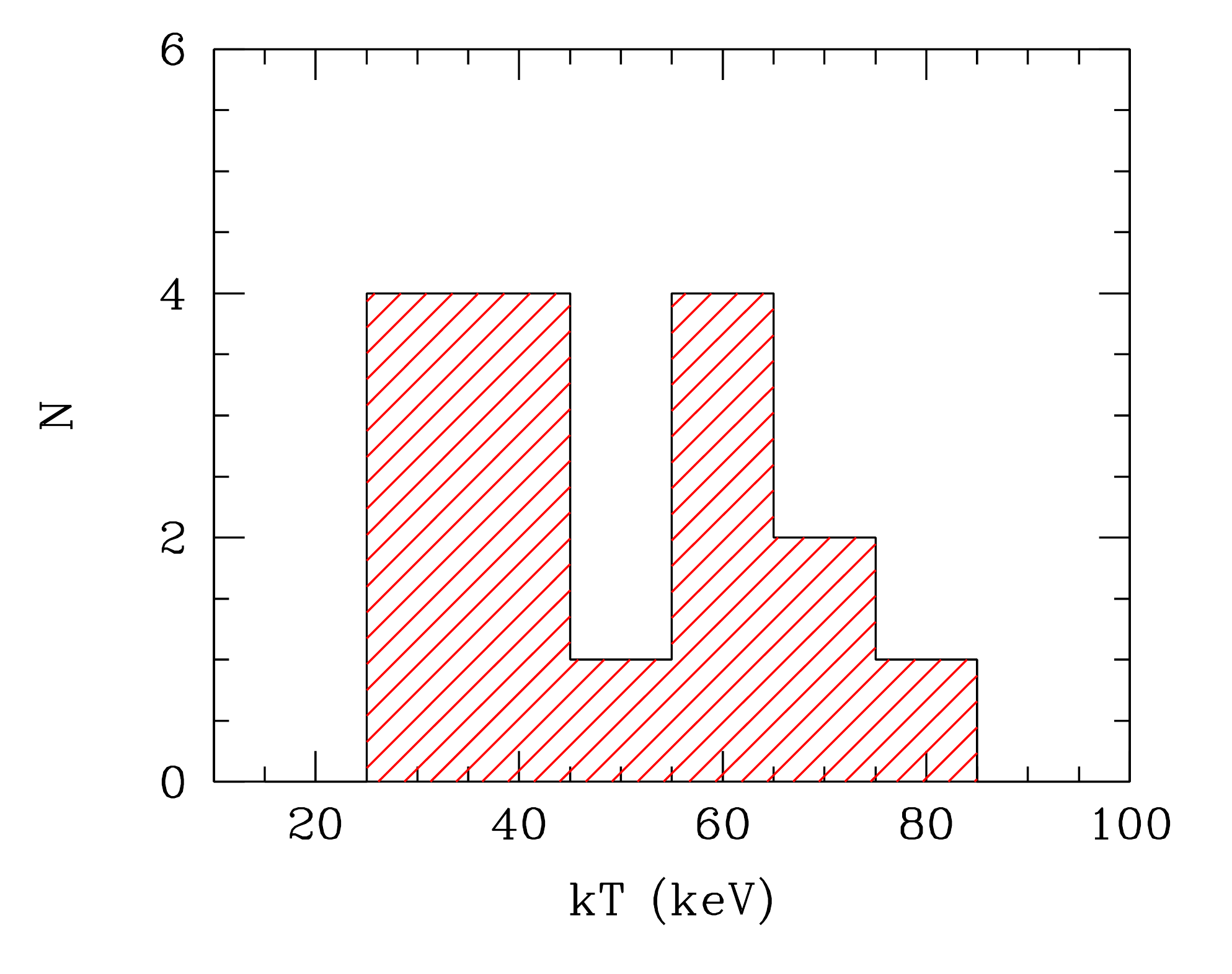}
\end{center}
\caption{The coronal temperature distribution as estimated using the 
{\sc COMPTT} model of \citet{titarchuk1994} (16 sources). The three sources in which kT is unconstrained have been excluded.}
\label{kT}
\end{figure}

\begin{table*}
\tiny
\caption{MYTORUS Spectral fits}
\centering
\setlength{\tabcolsep}{0.7mm}
\begin{tabular}{lcccccccc}
       \hline
Name & $\rm N_{H},Z$    &    $\rm kT$   & $\tau$         & $\rm E_C$ & $f^{refl}_{20-40keV}$& $\rm \chi^{2}/d.o.f. $  & Ref.\\
        (1) &   (2)                  & (3)         &    (4)           &       (5)       &          (6)                    &  (7)                      &   (8) \\     
\hline
NGC1068 &   $>10$     &   $67^{+97.0}_{-28.0}$    & $0.68^{+0.08}_{-0.23}$ &   $>$145  &     0.84 &  890/755 & a, i \\
 Circinus &   $9.0^{+1.00}_{-0.50}$     & $38^{+9.5}_{-7.6}$    & $0.61^{+0.10}_{-0.08}$ & $43^{+5}_{-4}$ &   0.80 & 1872/1724 & b \\
NGC6240 &  $1.3_{-0.22}^{+0.12}$ & $60_{-20.6}^{+85.5}$  & $1.7_{-0.73}^{+0.23}$& unconstr.  &  $<0.01$ & 456/496 & c, i\\  
NGC4945 &  $3.5_{-0.13}^{+0.20}$ & $59_{-13.3}^{+19.7}$ &  $1.7_{-0.32}^{+0.34}$  &   $>$42    & $ 0.02 $  &  1786/1743 & d, i \\
NGC424   & $1.3_{-0.20}^{+0.25}$  & $30_{-12.1}^{+20.9}$   & 2           & $>$64             &  0.93        &  147/121 & e, i\\
2MFGC2280 & $1.6_{-0.45}^{+0.72}$ &  50                  &   $1.3_{-1.3}^{+1.4}$  &    $<$238   &   0.03       &  66/59   & i \\
NGC1194     & $1.7_{-0.25}^{+0.24}$  & $37_{-1.0}^{+4.7}$  & $1.5_{-0.13}^{+0.30}$  & unconstr.&   $<0.01$  &  257/273 & f \\
MCG+06-16-028 & $1.5_{0.27}^{+0.42}$  & $34_{-18.1}^{+20.0}$  & $1.4^{+0.7}_{-1.3}$ & $>23$ & 0.04      & 136/132 & i \\
NGC3079            & $3.2_{-0.43}^{+0.54}$  & $31_{-8.1}^{+8.6}$     & 2  &   $56^{+354}_{-24}$    & 0.05       & 192/165   &  i \\
NGC3393            & $2.1_{-0.28}^{+0.61} $ &  $69_{-60.9}^{+77.9}$ &    $1.05_{-1.05}^{+0.97}$ & $>96$  &  $<0.01$ & 123/117 & g, i \\
NGC4941           & $1.2_{-0.39}^{+0.91}$    & $47^{+216}_{-34}$                        & 2 & unconstr. & 0.19 &  98/111 & \\
NGC5728           & $1.3_{-0.08}^{+0.13}$ & $81_{-30.2}^{+48.6}$ &   $1.5_{-0.41}^{+0.32} $ & $121^{+11}_{-13}$  &  $<0.01$&  444/436 & i \\
ESO137-34        & $2.9_{-0.54}^{+0.86}$  & $26_{-11.8}^{+15.9} $ & 2  & $>$203 &  0.12 & 175/147 & \\
NGC7212            & $1.1_{-0.27}^{+0.48}$ &  $41^{+32.2}_{-22.6}$                       & 2  &  $>56$ & $<$0.01 &  147/117 & i \\
NGC1229            &$0.6_{-0.14}^{+0.26}  $  & $60^{+51.2}_{-22.5}$                        & 2   & unconstr. &  $<0.01$ &  134/138 & i \\
NGC6232            & $0.8_{-0.30}^{+1.26}$ & $42_{-29.5}^{+40.2} $ & 2  & $>$500& $<0.01$ &   52/35 & i \\
2MASXJ09235371 & $1.0_{-0.18}^{+0.21}$ & $60_{-27.9}^{+68.1}$ & 2 &$35^{+6.1}_{-5.1}$ & $<0.01$ &  266/209 & i \\
NGC5643              & $1.3^{+0.10}_{-0.10}$  & 50                            &  2  &  $>$48 &  0.92      &  136/127 & h,i \\
NGC7130               & $2.3_{-0.31}^{+0.29} $ & 50 &2  & unconstr. & 0.23 & 116/126 & i \\
\hline
\label{spectral-fits}
\end{tabular}
\begin{itemize}
\item[] (1) Source name (2) Average column density ($N_H,Z=N_H,S90=N_H,S0$) in units of $\rm 10^{24}cm^{-2}$  (3) coronal temperature (keV); values with no error bar
were unconstrained and have been fixed to a value of 50 keV. 
(4) optical depth; values with no error bars were unconstrained and have been  fixed to a value of $\rm \tau=2$. 
 (5) the high energy cut-off derived by \citet{ricci2017} from {\it Niel Gehrels Swift, XMM-Newton, Chandra} joint fits. (6) unobscured reflected fraction in the 20-40 keV band. The unobscured reflection fraction is defined as $\rm Flux[S0]/(Flux[Z]+Flux[S90]+Flux[S0])$ (8) $\chi^2$ over degrees of freedom (9) Reference for previous {\it NuSTAR} \, analyses:
 (a) \citet{bauer2015};  (b) \citet{arevalo2014} ; (c) \citet{puccetti2014};  (d) \citet{puccetti2016}; (e) \citet{balokovic2014};  (f) \citet{masini2016} (g) \citet{koss2015}
 (h) \citet{annuar2015} (i) \citet{marchesi2018}. 
\end{itemize}
\end{table*}

\begin{figure*}
\begin{center}
\includegraphics[height=1.\columnwidth]{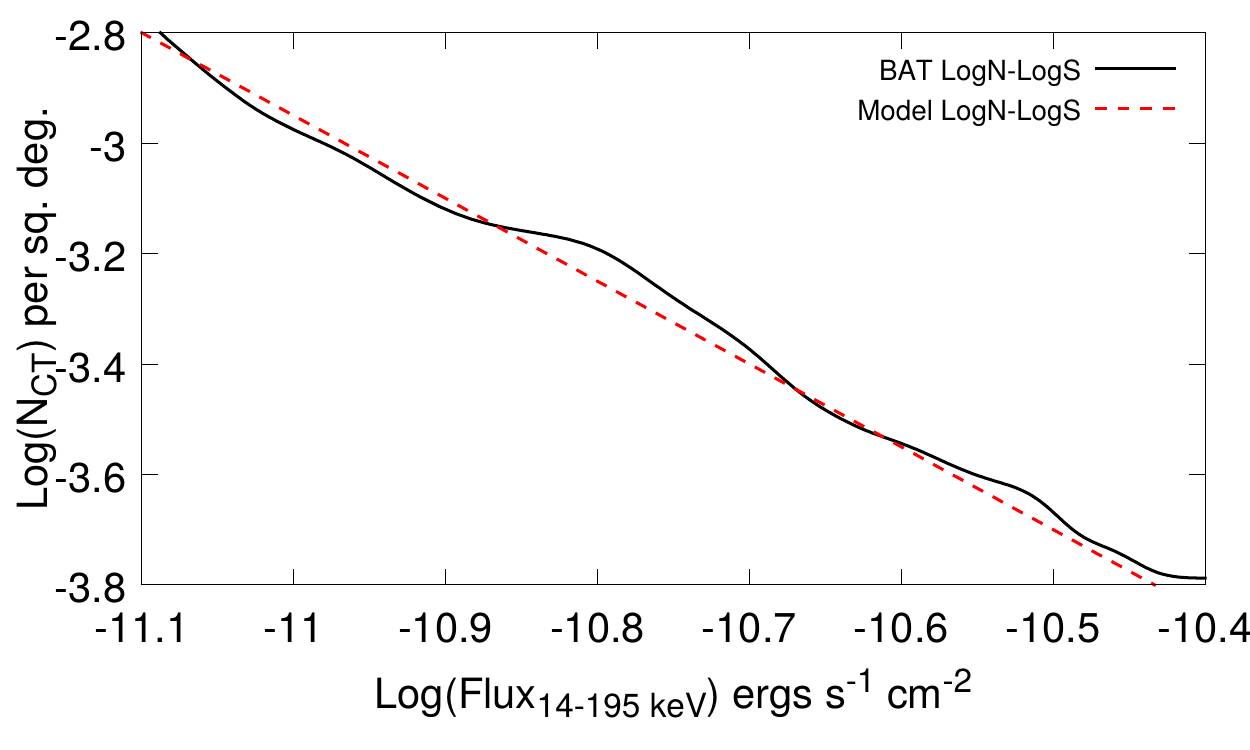}
\end{center}
\caption{The number count distribution of the Compton-thick AGN derived from the {\it Niel Gehrels SWIFT}-BAT survey in \citet{akylas2016} (black solid line). 
The red dashed  line denotes the prediction of the X-ray population synthesis model of \citet{akylas2012}. Using the spectral parameters  found in our spectral fits, 
we find that the Compton-thick AGN fraction should be 20\% of the type-2 AGN population, in order to agree with the observed logN-logS.}
\label{logNlogS}
\end{figure*}

\begin{figure*}[h]
\centering
\includegraphics[width=10.0cm]{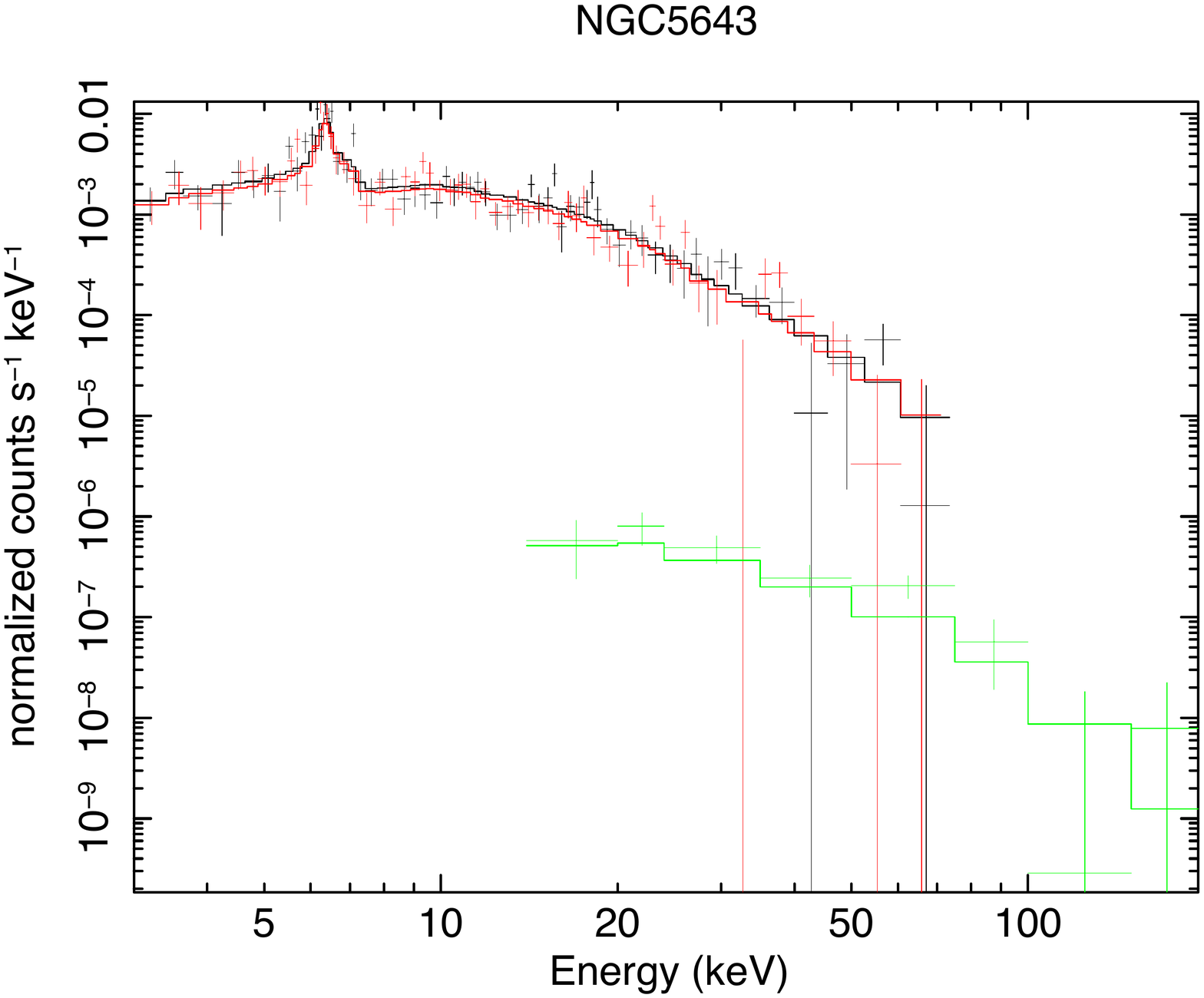}
\includegraphics[width=10.0cm]{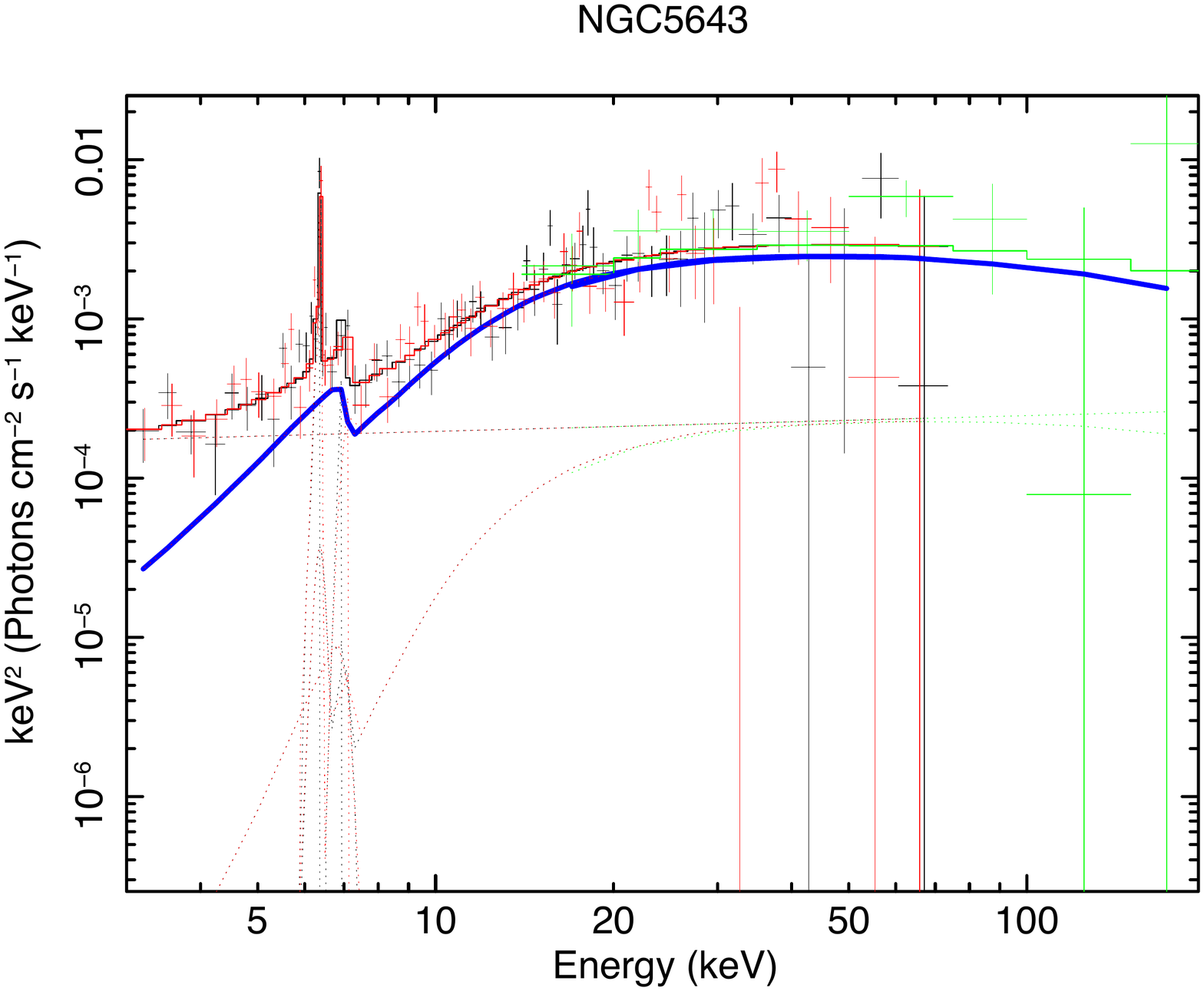} \\
\caption{NGC5643: a characteristic spectrum of an AGN with a high reflection component. Upper panel:  folded spectrum. Bottom panel:  unfolded spectrum; 
the blue sold line denotes the unabsorbed component S0. The straight dotted line represents the soft power-law component while the other dotted line denotes the 
absorbed reflection component S90. In both panels the black and red colours denote the two FPM {\it NuSTAR} detectors while the green colour corresponds to the BAT detector.
 }
\label{spectra}
\end{figure*}

\section{Results}
The spectral-fitting results are given in table \ref{spectral-fits}. In the following subsections we discuss separately our findings  regarding the hydrogen column density,
 the high energy cut-off and the reflection fraction. Finally, we discuss the implications of our spectral modelling for the intrinsic  number of Compton-thick AGN.  
 
 \subsection{Absorbing column density}
 The distribution of the column densities is presented in Fig. 1. The column densities found by {\it NuSTAR} are in very good agreement 
 with those previously found by {\it Niel Gehrels Swift}
\citep{akylas2016}. Only for one source (NGC7212), \citet{akylas2016} find a much higher column density.  The current {\it NuSTAR} results confirm 
the lower column density reported by \citet{ricci2015}. This difference is most probably because the {\it Niel Gehrels Swift}/XRT data have limited photon statistics while \citet{ricci2015} 
have combined the BAT data  with {\it XMM-Newton} spectra obtaining better quality photon statistics at energies below 10 keV.  
Recently, \citet{marchesi2018} 
published a {\it NuSTAR} analysis of Compton-thick AGN from the {\it Niel Gehrels SWIFT}/BAT list of \citet{ricci2015}. \citet{marchesi2018} find that a number of candidate Compton-thick AGN 
in \citet{ricci2015} may present lower column densities. There are 15 common objects between our sample and that of \citet{marchesi2018}. According to these authors four objects appear to have column densities somewhat lower than $\rm 10^{24} cm^{-2}$. These are NGC1194, NGC6232,  NGC1229 and 2MASSXJ09235371-3141305. According to our {\it NuSTAR} analysis, NGC1194 is above the Compton-thick threshold while NGC1229 and  2MASSXJ09235371-3141305 are 
consistent with being Compton-thick within the errors. In the case of NGC1229, although the source is unambiguously heavily obscured
 ($\rm N_H\approx 6^{+2.6}_{-1.4}\times 10^{23}cm^{-2}$ ) the column density derived   lies well below the Compton-thick limit.
For many other sources, e.g. NGC1068, Circinus, NGC424, NGC3079, NGC4945, NGC5643, {\it NuSTAR} spectral fits have been  
presented elsewhere \citep{bauer2015, arevalo2014, balokovic2014, puccetti2016, koss2015, puccetti2014, annuar2015, marchesi2018}. 
In general, our results are consistent with the work cited above. 

 As it can be seen from Fig. \ref{nh}, there is a clear tendency to detect  mildly obscured Compton-thick AGN with column densities  a few times $\rm 10^{24}cm^{-2}$.
This has been first noticed by \citet{burlon2011} who showed that the BAT instrument is unbiased only up to column densities  of only $\rm 2\times 10^{24} cm^{-2}$.
 We observe only two heavily Compton-thick source with a column density around $\rm 10^{25}cm^{-2}$. 
 Owing to their diminished X-ray emission, these heavily obscured sources can be detected only if they are relatively nearby. A similar effect is observed in 
 {\it INTEGRAL} observations of Compton-thick AGN 
 \citep{malizia2009}. Although the number of Compton-thick AGN is relatively limited in the current sample, 
  \citet{akylas2016} using a much larger sample noticed that the 
 column density distribution  is fully consistent with a flat Compton-thick column density distribution
 between $10^{24}$ and $10^{26}$ $\rm cm^{-2}$. 
 
\subsection{High energy cut-off}
 In column (3) of table \ref{spectral-fits} we give the derived coronal temperature. The cut-off energy corresponds to $\sim3kT$ \citep{titarchuk1994}. 
 For three objects the coronal temperature could not be constrained (2MFGC2280, NGC5643, NGC7130). 
 The mean energy of the coronal temperature of the remaining 16 sources is $\rm kT\approx  50$ keV corresponding to 
  a high energy cut-off of about $\rm E_C\sim 150$ keV.  The distribution of the coronal temperature is given in Fig. \ref{kT}. 
  In the case of the three sources above where the coronal temperature could not be well constrained, we fixed the temperature at 50 keV in order to 
  estimate the reflected emission. 
 
 It is interesting to compare the estimated cut-off with other high energy spectral studies of Compton-thick AGN. 
 \citet{dadina2007} estimated the cut-off energy in a number of nearby AGN using data from the {\it BeppoSAX} mission  which observes  in the 2-100 keV range. 
  There are eight sources in common between our sample and that of \citet{dadina2007}. Owing to the limited photon statistics at high energies, \citet{dadina2007} 
  were able to constrain the high-energy cut-off in only  two of these sources;
  NGC1068, with $\rm E_C>116$ keV and NGC4945 with $\rm E_C=122^{+41}_{-26}$ keV. These values are entirely consistent with our combined {\it NuSTAR} and BAT analysis. 
  \citet{ricci2017} derived the cut-off energy of the Compton-thick AGN in his sample using a combination of {\it XMM-Newton} , {\it Chandra}, and {\it Niel Gehrels Swift} spectra
   (both XRT and BAT). For comparison with the present work, the cut-off energies, $\rm E_C\sim3kT$, derived by \citet{ricci2017},  are given in column  (5) of table \ref{spectral-fits}. 
   In many cases, there is reasonable agreement between the two analyses.  However, there are at least four cases where the results are clearly at odds. 
    These are Circinus, ESO137-34, NGC6232 and 2MASXJ09235371. This discrepancy could be partly attributed to the different spectral models used that is 
    power-law model with an exponential cut-off versus the comptonization models. \citet{ricci2017} have determined the average cut-off energy for their Compton-thick sources. 
   When they exclude the lower and upper limits, they   find a very low high energy cut-off for their Compton-thick AGN ($\rm E_C \approx 43\pm 15$ keV). 
   Taking lower and upper limits into account by means of the Kaplan-Meier estimator, they estimate a much higher value for the  cut-off 
    $\rm E_C\sim 450\pm 64$ keV. 
   We find a mean high energy cut-off of $\rm E_C\sim3kT\approx150$ keV, excluding the three sources above where the cut-off energy could not be constrained.
          It is also interesting to compare with the work of  \citet{dadina2007} who derived the average X-ray spectrum of 42 type-2 AGN using {\it BeppoSAX} data. 
   They find a cut-off energy of $\rm E_C=376\pm42$ keV. This is significantly higher than the cut-off energy of our Compton-thick sources. 

Our results are not significantly affected by the choice of $\tau=2.0$  in the cases where the photon statistics is limited.
For example, in the case of NGC3079 if we choose $\tau=1.5$, the resulting temperature  becomes $\rm kT= 39\pm19$ keV. For this temperature of 
the corresponding photon index would be $\Gamma\approx2.0$.

\subsection{Reflection component}

 The fraction of the reflected emission in the 20-40 keV band plays an important role in the prediction of the intrinsic Compton-thick AGN fraction, 
  in the sense that as  the reflected emission 
becomes higher, the number of Compton-thick AGN required is reduced. This is because the higher the reflection the easier is to detect Compton-thick 
AGN in the BAT band and hence the number of 'missing' Compton-thick AGN is relatively small. The reflection fraction is given in column (6) of table 2. 
This is defined as the ratio of the
unabsorbed reflection component S0 versus the total flux i.e. the sum of the line of sight component Z and the two scattered components S0 and S90. 
 Four sources appear to be reflection dominated as the reflection radiation fraction is near unity. These are NGC1068, Circinus, NGC424 and NGC5643. 
 A typical spectrum of a source with a strong reflection component (NGC5643) is shown in Fig. \ref{spectra}. 
There is not a clear correlation between the strength of the reflection component and the column density in these sources. 
For example, NGC424 and NGC5643 have 'moderate' Compton-thick  column densities 1.3 $\rm \times10^{24} cm^{-2}$. 
The vast majority of the sources have moderate strength reflection components below 0.23.
 There are eight  sources where the best-fit reflection component is negligible i.e. well below 0.01.  The distribution of the reflection 
fraction is given in Fig. \ref{refl}. The mean reflection  fraction is 0.22.  In analogy to the commonly used  {\sc PEXRAV} model \citep{Zdziarski1995}, 
which estimates the reflection  to an infinite slab, we find that a fraction of 0.22 corresponds to a reflection strength R$\approx$1. 
This is equivalent to  a reflection fraction of $\sim$3\% in the 2-10 keV band. 

\subsection{The intrinsic fraction  of Compton-thick AGN}
In this section we estimate the intrinsic number of Compton-thick AGN in the Universe using the population synthesis models developed by 
\citet{akylas2012}. The X-ray background synthesis models have been primarily being devised in order to predict the number of 
Compton-thick AGN based on their contribution to the X-ray background spectrum and in particular at its energy peak around 20-30 keV. 
The X-ray background synthesis models predict the number of AGN and their contribution to the X-ray background spectrum 
at any redshift, luminosity and column density. The necessary ingredients are the AGN luminosity function and the typical AGN spectrum. 
\citet{treister2009} first pointed out that uncertainties on the measurements of the X-ray background are substantially large to hamper an accurate prediction 
of the number of Compton-thick AGN. Instead they propose that the  {\it INTEGRAL}  or BAT number counts of Compton-thick AGN place tighter constraints. 
 In this manner,  the X-ray background synthesis models could be used to extrapolate the observed number counts to the intrinsic population number counts. 
 In a similar approach, \citet{burlon2011} use the observed Compton-thick AGN number counts and  predict the intrinsic Compton-thick AGN counts by 
 assuming a relation between the observed and the intrinsic 
 Compton-thick flux. This conversion sensitively depends on the typical Compton-thick AGN spectrum. Using a simple spectrum 
 and by deriving the intrinsic $\rm N_H$ distribution, \citet{burlon2011} derive an intrinsic Compton-thick AGN contribution of $\sim$20\%.
   
   Here, we fold the full observed spectral distribution  as described by the 
   parameters $\rm N_H$, kT and the reflection fraction (see Fig.1, Fig.2 and Fig.3) in our X-ray population synthesis models. 
   Then,  the remaining unknown parameter is the intrinsic fraction of Compton-thick AGN.
 We find that in order to match the observed Compton-thick AGN counts
  in the BAT 14-195 keV band with our models, we need an intrinsic 
 fraction of $\sim$20\% relative to the type-II AGN population ($\rm N_H>10^{22} cm^{-2}$), see Fig. \ref{logNlogS}. 
 This corresponds to a Compton-thick fraction of 15\% relative to the total AGN population.
 The statistical error on this figure is 3\%. This is   derived from the 
 number of Compton-thick AGN ($53\pm7$) used to  derive the logN-logS. 
 The uncertainties on the spectral parameters estimated above have not been taken into account in this error estimate. 
 
 Recently, \citet{masini2018} estimated the Compton-think fraction of Compton-thick sources, using {\it NuSTAR} 
 observations in the UKIDSS Ultra deep field. They find an observed Compton-thick fraction of $8.4\pm2 \%$ 
 in the 8-24 keV band at a flux of $2.7\times 10^{-14}$ $\rm erg~cm^{-2}~s^{-1}$. Making corrections for their non-detected  {\it XMM-Newton}
 sources this fraction becomes $11.5\pm2.0$. This is in agreement with the 1$\sigma$ upper limit of our model 
 predictions which is $\approx10\%$  at this flux.

\section{Summary and Conclusions}
Direct spectroscopic observations of AGN in the local Universe, using {\it Niel Gehrels Swift}/BAT samples, find a fraction of Compton-thick AGN of about 
7\% \citep{burlon2011, ricci2015, akylas2016, marchesi2018}. This figure can be extrapolated to the intrinsic number of Compton-thick AGN 
by assuming a spectral model for the Compton-thick AGN spectrum \citep[e.g.][]{burlon2011}. The spectral parameters of interest are 
the photon index, the strength of the reflection component, the column density distribution, and the primary emission cut-off energy. 
  The reflection of material on cold matter surrounding the nucleus results in the flattening of the spectrum around 20-30 keV. 
 Therefore the  higher the reflection the easier is to detect Compton-thick AGN in the 14-195 keV band and hence the 
 lower  the number of missing  Compton-thick AGN (the lower the intrinsic fraction). The high energy cut-off works in the opposite way.  
 The lower the value of the cut-off energy, the higher the intrinsic 
 number of Compton-thick AGN  as these would not be easily detected in the BAT band.  
 For example, if the  high energy cut-off were $\approx$50keV 
 \citep[see e.g.][]{ricci2017}, the fraction of Compton-thick AGN fraction would increase to about 50\%. 
 Note that in contrary, cut-offs higher than 150 keV would not  significantly affect our results. 

 Aiming to constrain the  intrinsic fraction of Compton-thick AGN in the local  Universe, we explore the spectrum of a large number
   Compton-thick AGN. We analyse the combined {\it NuSTAR}/BAT spectra of 19 candidate Compton-thick AGN (at the 70\% confidence level)
  in the sample of\citet{akylas2016}. We model the spectra using the {\sc MYTORUS} models of \citet{yaqoob2012} which take 
  into account absorption and reflection from Compton-thick material.  
  The primary X-ray emission from a hot corona is modelled with a  physically motivated model 
  \citep{titarchuk1994}. Our results can be briefly summarised as follows.
  
  \begin{itemize}
  \item  We derive absorbing column densities which are consistent with those derived by the previous {\it Niel Gehrels Swift}/BAT analyses
  by \citet{ricci2015} and \citet{akylas2016} .
  
  \item  We estimate that the coronal temperatures  lie between 25 and 80 keV corresponding to high energy cut-offs
 roughly between 75 and 240 keV. The mean cut-off energy is 150 keV. 
  
 \item We find that a large fraction  of our AGN lack a significant reflection component 
 in the 20-40 keV band. The average reflection fraction is $\sim$0.22 . 
 
  \item The resulting spectral parameters are used to place constraints on the intrinsic Compton-thick 
 AGN fraction. Using the above spectral results  in combination with our synthesis models,
  we extrapolate the observed fraction of Compton-thick AGN to the intrinsic fraction. 
  We find that the Compton-thick AGN fraction is $20\pm3$\% of the type-II AGN population. 
  \end{itemize}

\begin{acknowledgements}
We would like to thank the anonymous referee for comments and suggestions which helped to improve the paper. 
We acknowledge the use of {\sc MYTORUS} spectral models by \citet{yaqoob2011}. We would also like to thank Tahir Yaqoob for his help and suggestions 
in the application of the above models. We would like to thank the anonymous referee for many comments and suggestions which helped to improve 
substantially this paper.
The research leading to these results has received funding from the European Union's Horizon 2020 
Programme under the AHEAD project (grant agreement n. 654215). 
\end{acknowledgements}

\bibliography{ref.bib}{}
\bibliographystyle{aa}

\end{document}